\documentclass[]{spie}  
 
\usepackage{amsmath,amsfonts,amssymb}
\usepackage{graphicx}
\usepackage[symbol]{footmisc}
\usepackage[colorlinks=true, allcolors=blue]{hyperref}
\graphicspath{{./Figures/}}

\title{CHARA/Silmaril Instrument Software and Data Reduction Pipeline: Characterization of the Instrument in the Lab and On-Sky}

\author[a]{Narsireddy Anugu}
\author[a]{Theo A. ten brummelaar}
\author[a]{Cyprien Lanthermann}
\author[b]{Peter G. Tuthill}
\author[a]{Edgar R. Ligon III}
\author[a]{Gail H. Schaefer}
\author[a]{Douglas R. Gies}
\author[b]{Grace Piroscia}
\author[b]{Adam Taras}
\author[c]{Gerard T. van Belle}
\author[d]{Makoto Kishimoto}
\author[b]{Marc-Antoine Martinod}

\affil[a]{The CHARA Array of Georgia State University, Mount Wilson Observatory, Mount Wilson, CA 91023, USA}
\affil[b]{Sydney Astrophotonics Instrumentation Laboratory, School of Physics, The University of Sydney, Sydney, NSW, 2006, Australia}
\affil[c]{Lowell Observatory, 1400 West Mars Hill Road, Flagstaff, AZ 86001, USA}
\affil[d]{Department of Astrophysics \& Atmospheric Sciences, Kyoto Sangyo University, Kamigamo-motoyama, Kita-ku, Kyoto 603-8555, Japan}

\authorinfo{Send correspondence to Narsireddy Anugu: E-mail: nanugu@gsu.edu}

\begin{document} 
\maketitle

\begin{abstract}
The newly installed Silmaril beam combiner at the CHARA array is designed to observe previously inaccessible faint targets, including Active Galactic Nuclei and T-Tauri Young Stellar Objects. Silmaril leverages cutting-edge optical design, low readout noise, and a high-speed C-RED1 camera to realize its sensitivity objectives. In this presentation, we offer a comprehensive overview of the instrument's software, which manages critical functions, including camera data acquisition, fringe tracking, automatic instrument alignment, and observing interfaces, all aimed at optimizing on-sky data collection. Additionally, we offer an outline of the data reduction pipeline, responsible for converting raw instrument data products into the final OIFITS used by the standard interferometry modeling software. Finally, a thorough analysis of the camera and instrument characterization results will be presented, evaluating instrument performance in terms of sensitivity. The purpose of this paper is to provide a solid reference for studies based on Silmaril data.
\end{abstract}


\section{Introduction}
Silmaril is a new instrument installed at the CHARA Array \cite{tenBrummelaar+2005}, designed to combine light from $2\times3$ telescopes operating in the near-infrared H and K bands simultaneously. The details of its optical design, simulations, and operational concepts are described in the 2022 and 2024 SPIE proceedings \cite{Lanthermann2022,Lanthermann2024}. We summarize the key instrument concepts here.  CHARA already has six telescope beam combiners operating at different wavelengths (R \cite{Mourard+2022}, JH \cite{Anugu+2020}, and K \cite{Setterholm2023JATIS...9b5006S} bands).  MIRC-X and MYSTIC can observe targets up to $H/K \lessapprox 7.5$. Silmaril aims to access targets much fainter than this limit. The detailed scientific goals of Silmaril are outlined in Lanthermann et al. (2022) \cite{Lanthermann2022}. In essence, Silmaril aims observations of previously inaccessible objects like Active Galactic Nuclei and faint young stellar objects (T-Tauri stars). The instrument sensitivity is achieved through a combination of cutting-edge design elements.  Silmaril prioritizes maximizing the amount of light collected,  over achieving higher spectral resolution. Unlike other CHARA combiners, it utilizes bulk optics instead of integrated optics and single-mode fibers to limit the light coupling losses. Furthermore, by minimizing the number of optical components in the light path, this design aims increased throughput efficiency. To minimize thermal (blackbody) background radiation, Silmaril employs a two-pronged approach: a very long focal length cylindrical mirror ($\sim5.4$~m) and a smaller cold stop aperture ($F/20$) inside the camera. Additionally, we are planning to implement a technique called ``Narcissus reflection" to further reduce blackbody radiation\cite{Lanthermann2024,Taras2024ApOpt..63D..41T}. This technique blocks unwanted radiation from high-emissivity areas within the instrument, preventing it from reaching the detector. Finally, to minimize readout noise, Silmaril leverages the proven C-RED1 detector system used in MIRC-X \cite{Anugu+2020} and MYSTIC \cite{Setterholm2023JATIS...9b5006S}.

Silmaril was installed at the CHARA lab in early 2023. By June 2023, we successfully achieved first light fringes on-sky using an engineering C-RED2 camera. We received the science-ready C-RED1 camera in September 2023. However, our laboratory characterization revealed camera performance issues, necessitating its return for repairs. We are currently awaiting for the repaired C-RED1 camera to resume on-sky fringe observations.

This paper serves as a user manual for Silmaril observing users, guiding them through its operation, data collection, and data reduction software. 

\section{Technical Requirements}
\label{sec:tech-rec}

In this section, we delve into the technical requirements of the instrument, focusing on aspects crucial for operation, data collection, and data reduction.

\begin{figure}
\centering
\includegraphics[width=\textwidth]{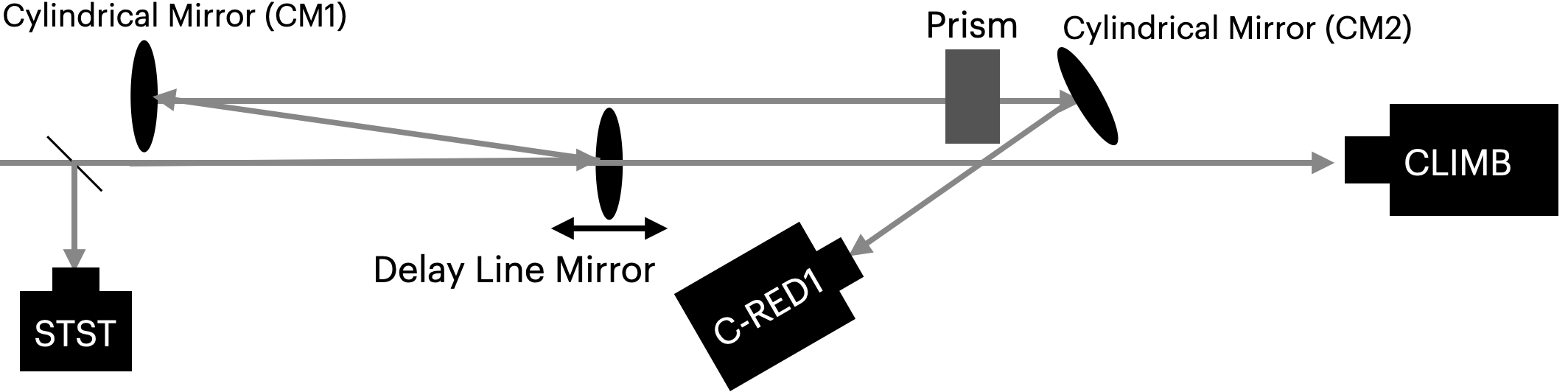}
\caption 
{\label{fig:concept}
Conceptual sketch of Silmaril -- with one beam. The beams travel from left to right. Initially, the CHARA beams are redirected by the Silmaril internal delay line mirrors to the long-focal Cylindrical Mirrors (CM1). CM1 has a focal length of $5.4$~m, which focuses the beam along the spatial (fringe) direction on the detector. The Prism, with a spectral resolution of $R \sim 35$, disperses the light. CM2 has a focal length of 350 mm, focus the beam on the detector along the spectral direction. The star trackers, STST and CLIMB can be inserted into the optical path or moved out. The delay line mirror is moved by a Zaber linear motion actuator.} 
\end{figure} 

\begin{figure}
\centering
\includegraphics[width=\textwidth]{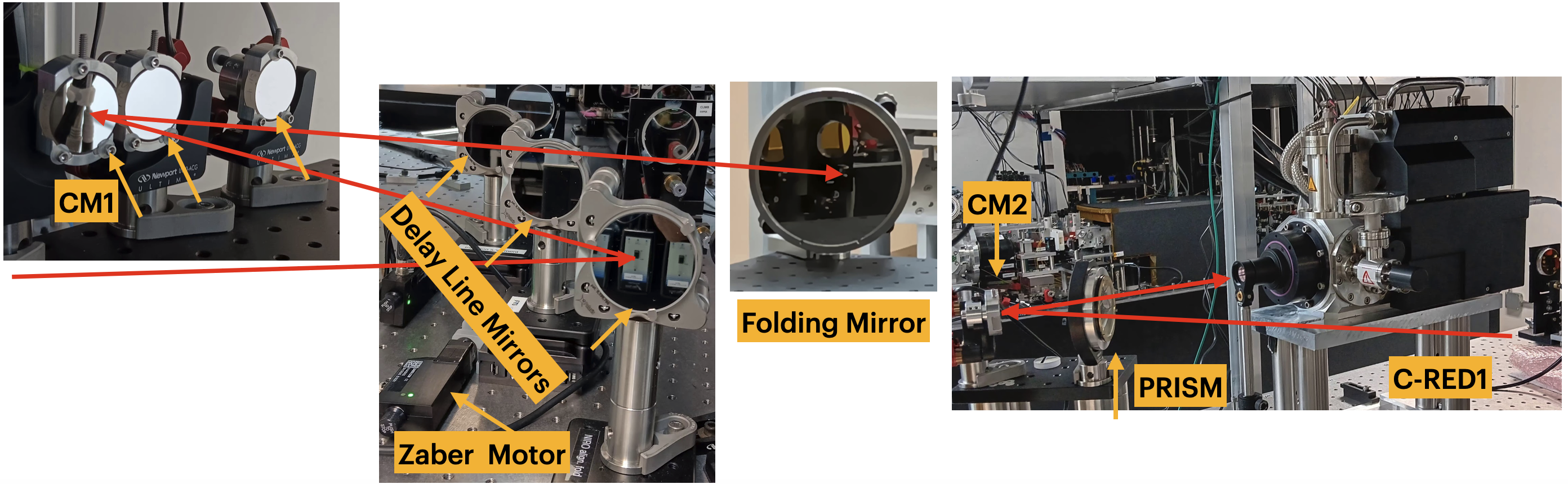}
\caption 
{\label{fig:optics} The installed optics and motor actuators of the Silmaril instrument are shown in Figure~\ref{fig:concept}. CM1 and CM2 are equipped with picomotors for the alignment. The delay line mirrors are equipped with Zaber motors, which move linearly and introduce optical path delay.  } 
\end{figure}

\begin{enumerate}
    \item Detector Data Acquisition: The data acquisition system should support frame rates up to 1~kHz for fringe tracking to stabilize fringes against atmospheric turbulence.
    \item Beam Combination:  The Silmaril instrument currently combines 3 beams but has been planned with the capability to expand to combine $2 \times 3$ beams separately. The two separate combined fringes, referred to as the left and right combinations, are recorded in different region of interest windows on the detector. The software must provide a separate server for each beam combination to read the fringe windows, perform Fourier transforms, compute group delay tracking, stabilize fringes by correcting atmospheric optical path length delays, and save the raw data in FITS format.
    \item Software Configuration: The same software should be used for both servers (left and right beam combiners), with configurations specified by startup options such as \texttt{silmaril\_server --left} and \texttt{silmaril\_server --right}. The Graphical User Interfaces (GUI) can be  started such as \texttt{silmaril\_rtd\_gtk --left} or \texttt{silmaril\_rtd\_gtk --right}. 
    \item Beam Stabilization: Our initial on-sky results indicate that we experience slow beam drifts, which can affect the observations in two ways: (i) imperfect beam overlap, leading to biased measurements of fringe contrast, which leads to large error bars in the squared visibility ($V^2$) and closure phase ($CP$); and (ii) reduced fringe signal-to-noise ratio (SNR) data, affecting the instrument sensitivity. To address these issues, the software must provide active beam stabilization. This can be achieved by measuring slow tip-tilts using the star tip-tilt tracker (either the Six Telescope Star Tracker, STST \cite{Setterholm2023JATIS...9b5006S} or CLIMB \cite{tenBrummelaar+2013}) and and correcting them with the beacon flat mirror located on the telescope adaptive optics bench and the M7 mirror in the telescope path.  The STST can currently stabilize beams down to $J/H \sim6$ mag and that we are planning to use the CLIMB camera to extend the correction to fainter stars. Figure~\ref{fig:concept} presents a conceptual sketch of the Silmaril instrument, including the STST and CLIMB components. Figure~\ref{fig:optics} shows the as-built optical  and their opto-mechanical components implementation.  The STST and CLIMB components receive light by inserting beam splitters into the optical path using motorized stages. The STST is already installed and utilizes approximately 5\% of the light for tip-tilt tracking and photometry collection, which can also be used for the data reduction pipeline. The amount of light to be shared with CLIMB is yet to be determined. Since we aim to correct only slow beam drifts (with frequencies of a few minutes), long-exposure tip-tilt measurements (tens of seconds to a minute) could suffice with a smaller percentage of flux split. Another option is to use a dichroic to send the entire J-band to CLIMB, resulting in a high SNR for tip-tilt tracking. However, the downside is that this photometric information could not be efficiently used for the data reduction pipeline because of wavelength differences to fringe acquisition (H and K). Furthermore, this dichroic option could be challenging for red colored objects, such as AGN and T-Tauri stars, which are fainter in the J-band compared to the K-band.
    \item Data Reduction Pipeline: A data reduction pipeline is required to process the datasets of  both $2\times3$ beam combiners.
    \item Automation: Silmaril must be a fully automated system to enable remote operations.
\end{enumerate}

\begin{figure}
\centering
\includegraphics[width=0.8\textwidth]{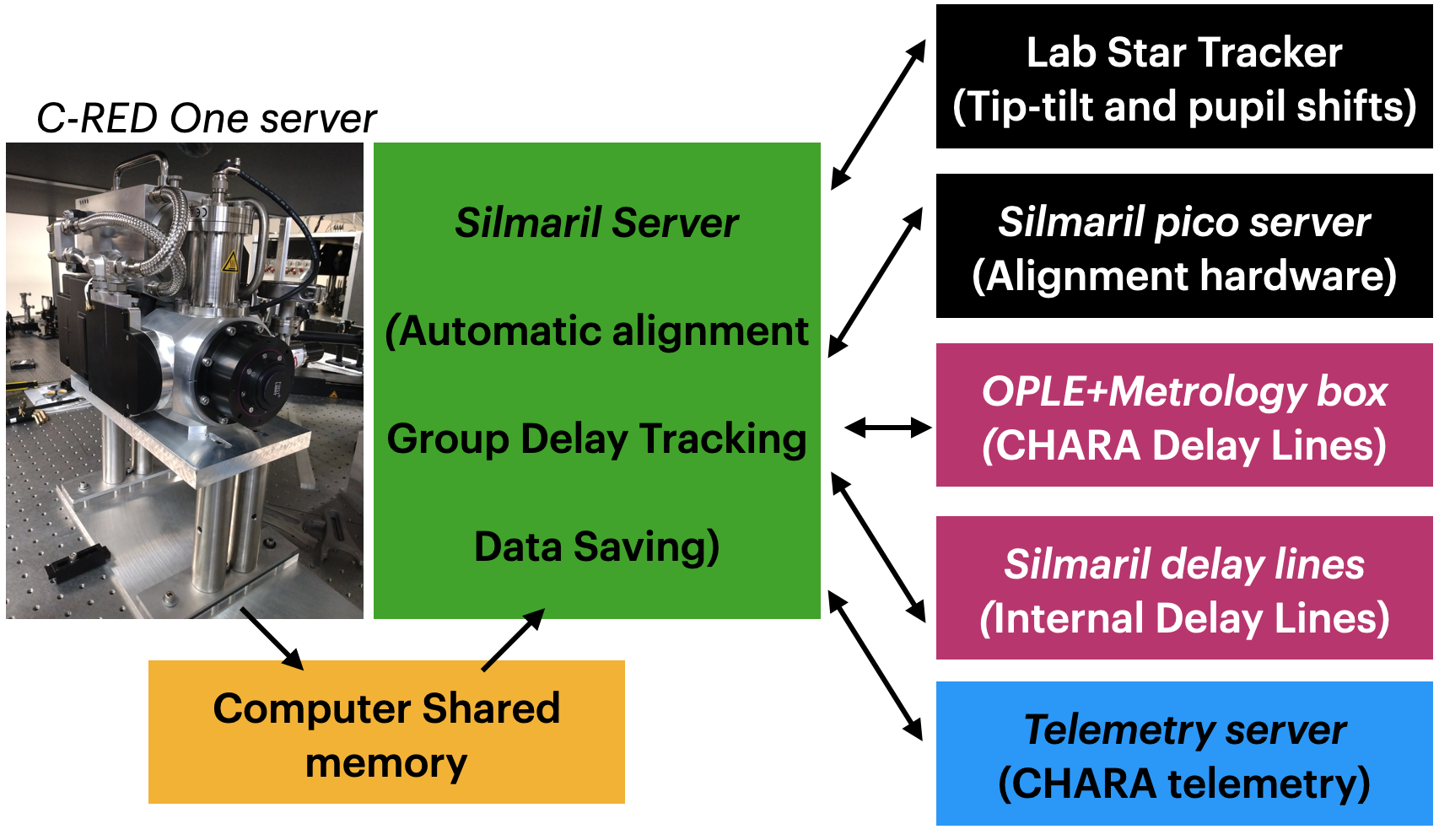}
\caption 
{\label{fig:server_software} The C-RED1 frames are read by the \texttt{C-RED One server} and written to a computer shared memory. These frames are then read by the \texttt{silmaril\_server}, which implements the core functionalities of the instrument. These functionalities can be divided into three tasks: (i) (Auto)alignments; (ii) Data analysis, including Fast Fourier Transforms (FFTs) and computation of group delay tracking; (iii) Data saving. The alignments are performed based on the Lab Star Tracker tip-tilt measurements obtained from either the STST or CLIMB camera. The tip-tilt shifts either sent to the telescope/lab adaptive optics or to the pico actuators of CM1. The internal delay lines are used to achieve cophasing with the STS. The group delay control is implemented to stabilize the fringes by sending the measured atmospheric OPDs to the CHARA delay lines. } 
\end{figure}

\begin{figure}
\centering
\includegraphics[width=0.5\textwidth]{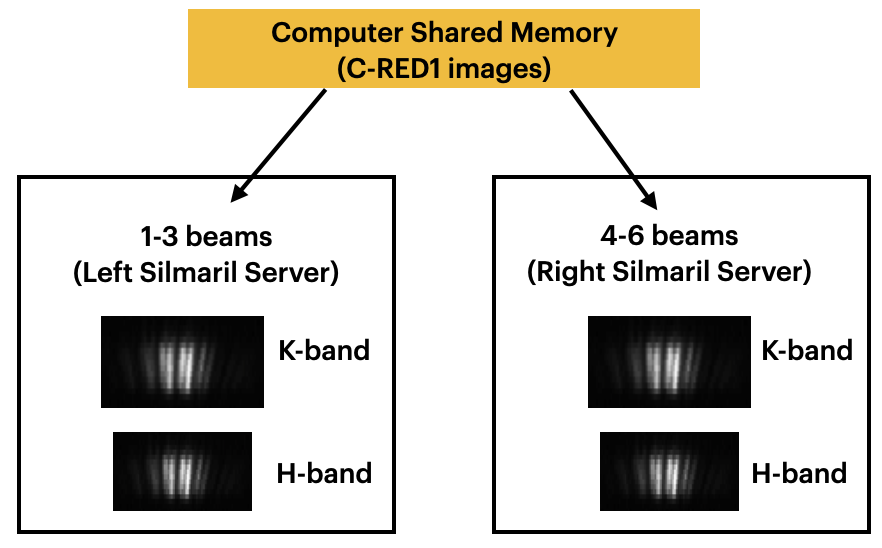}
\caption 
{\label{fig:super} A sketch of fringes of beams 1-3 and 4-6 are recorded on the same C-RED1 detector but in different sub-regions/windows. Each beam combination produces fringes in both H and K bands simultaneously. These sub-windows are read by the left and right \texttt{silmaril\_servers}, which perform all the core functions as described in Fig. \ref{fig:server_software}. The left and right \texttt{silmaril\_servers} work independently. For example, the left \texttt{silmaril\_server} controls the first three beams/telescopes, manages its own backgrounds, computes FFTs and group delay tracking, and sends the OPDs to its configured CHARA delay lines to stabilize the fringes.  } 
\end{figure}

\begin{figure}
\centering
\includegraphics[width=0.5\textwidth]{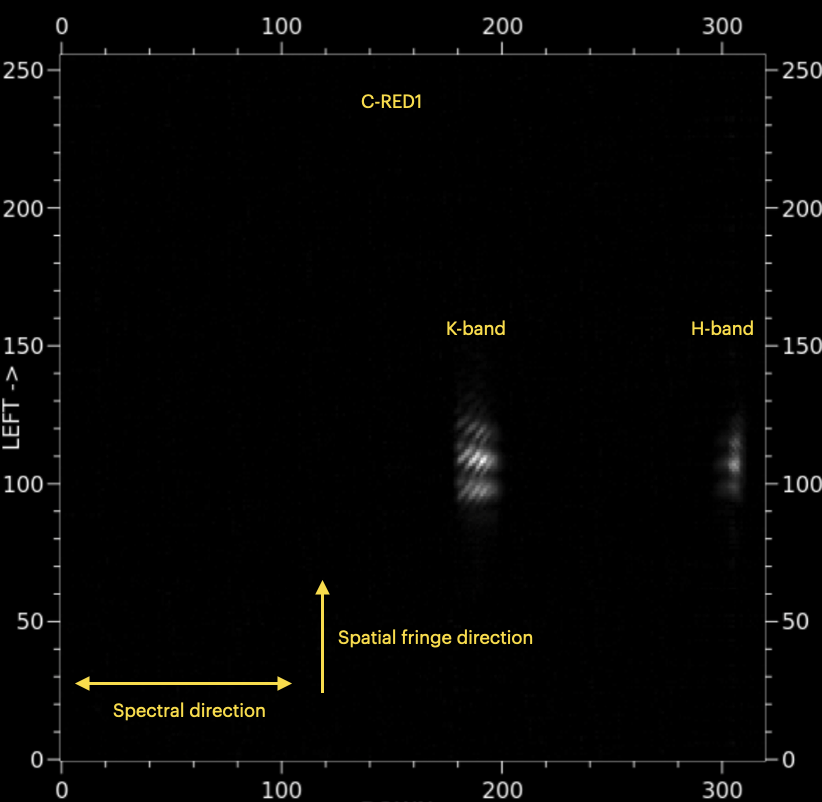}
\caption 
{\label{fig:cred1} \texttt{silmaril\_rtd\_gtk} displays the H and K-band fringes as recorded by the C-RED1 camera.  Silmaril incorporates an ``edge filter", which separates H and K band light while reducing background noise specifically in the H band channel\cite{Lanthermann2022}.} 
\end{figure}

\section{Real-Time Instrument Software and Operational GUIs}\label{Sec:software}

\subsection{Instrument software}

Fig~\ref{fig:server_software} presents the software architecture of the Silmaril instrument. The instrument software  manages critical functions of the instrument operation, including C-RED1 or C-RED2 camera data acquisition, fringe tracking, automatic instrument alignment with various actuator control, and observing interfaces, all aimed at optimizing on-sky data collection. The software design of Silmaril draws inspiration from the MIRC-X instrument for two main reasons.  Firstly, both instruments use similar camera and all-in-one beam combiner. This strategic decision leverages a proven design, saving time and effort compared to developing a completely new system (``reinventing the wheel"). Secondly, the MIRC-X codebase serves as a well-established standard practice for CHARA observations. By adopting a similar approach, Silmaril benefits from the inherent reliability and familiarity associated with the MIRC-X software. However, there is one key difference. Our Silmaril data acquisition software utilizes the newer C-RED1, software development kit (SDK, Ubuntu 16 version 2.95), specifically designed for the latest version of the Silmaril camera and firmware version. This was required because the 2017-era MIRC-X software  was not  compatible with more recent version of the camera. We use the similar computer as MIRC-X and runs Linux Xubuntu operating system. The code is written in C/C++ based on the CHARA server/client architecture. 

The Silmaril instrument software can be subdivided into three main parts.

\subsubsection{Camera Data Acquisition} Camera data acquisition is implemented by a server named \texttt{simaril\_credone}. This server performs frame grabbing at high speeds and with low latency. Once the images are read from the detector, they are written to a shared memory segment with low latency. These images are then accessed by \texttt{simaril\_server} for the image processing. The detector frames are required to read much faster than the atmospheric coherence time to stabilize the fringes. The atmospheric coherent time at the CHARA array ranges typically from 5-20 ms depending on average to good seeing nights. We adapted to the maximum frame rates (300 Hz) used in MIRC-X and MYSTIC. Another aspect of data acquisition is low latency. Higher latency or missing frames can cause delays, making corrections applied to delay lines ineffective since they are not synced with the atmospheric piston error. For the initial development of the instrument, we used C-RED2 while awaiting the final science version of the camera, C-RED1.

The data acquisition from the C-RED1 and C-RED2 are implemented with a Matrox Radient eV-CL frame grabber installed on the computer and connected to the camera via two camera link cables lengths of 10 m. C-RED1 uses the SAPHIRA detector featuring 32 parallel video channel outputs organized into 10 column blocks. Each channel output reads 32 adjacent pixels in a row simultaneously. With a pixel clock set to 10~MHz, the detector achieves a readout speed of approximately 640 Mpixels per second. This allows for reading a full $320\times256$ pixel frame at a rate of 3500 frames per second (FPS). The C-RED1 detector offers several readout modes, with the IOTA mode (``Fowler sampling") being particularly useful for further reducing the readout noise\cite{Anugu+2020}. In this mode, each pixel is read $N$ reads  times before moving to the next pixel in the row, and each row is read  $M$ loops  times before moving to the next row. This approach enables reduction in readout noise by approximately the $\sqrt{N\times M}$. Additionally, several frames are read nondestructively a predetermined number of times before the detector is reset, a process known as ``up the ramp" sampling. The final exposure image is then obtained by subtracting any two successive frames within the ramp. We chose to reset the frames per ramp for every 50 frames for fainter stars, although this number is shortened for brighter targets.

\subsubsection{Fringe Tracking in H/K-bands or Combined} The fringe tracking and locking to stabilize the fringes is implemented by  servers named \texttt{silmaril\_server --left} and \texttt{silmaril\_server --right}. These servers are the top level servers and read images from the shared memory written by \texttt{silmaril\_credone} and performs image processing on the appropriate fringe windows (see Fig~\ref{fig:super}). The raw images are first subtracted by backgrounds. Fourier transforms are then applied to these images to estimate the group delay optical path delay (OPD) offsets using the spectrally dispersed fringes. These OPD offsets are sent to the CHARA delay lines to stabilize the fringes against atmospheric turbulence and optical vibrations along the optical path.

Internal Zaber actuators mounted to the delay line mirror (see Fig~\ref{fig:concept} and \ref{fig:optics}) are used to co-phase with the Six Telescope Simulator (STS)\cite{Anugu+2020}, which serves as the reference for the CHARA astrometric fringe-finding baseline solution. The internal delay lines are moved with Zaber actuators, part number X-NA08A25, which have a range of 25.4~mm.

To utilize the simultaneous H and K-band fringes (see Fig~\ref{fig:cred1}), we implement two fringe tracking modes: (i) Primary mode and (ii) Combined mode. The Primary mode leverages the color-magnitude difference of targets in H and K bands, selecting the more sensitive band as the fringe tracker. In the Combined mode, we perform FFTs and group delay fringe tracking computation on both H and K-band fringe windows. This mode combines the correlated flux information from both bands, selecting the higher SNR weighted fringes (i.e., OPDs) for robust high-SNR fringe tracking, particularly useful for over-resolved objects.

Having fringes detected in both H and K-band wavelengths, the combined data allows to detect the smaller and extended features of an object. The angular resolution is given by $\lambda/2B$, where $\lambda$ is the wavelength and $B$ is the baseline separation between any two telescopes.

\subsubsection{Beam Alignment and Beam Stabilization} The alignment of the instrument is done in two steps, implemented inside the \texttt{silmaril\_server}. The first-order coarse alignment is achieved by moving the cylindrical mirrors (CM1s) internal to the Silmaril. The CM1 and CM2 are moved by Picomotor controller, 8742-8-KIT from Newport. This initial alignment is executed by moving pico motors mounted on the cylindrical mirrors. Once the flux is located, the second-order beam stabilization keeps the beams centered on the reference pixels on the detector using a tip-tilt tracker. For beam stabilization, we plan to use either STST or CLIMB six beam tip-tilt trackers. This tip-tilt signal can then be used to adjust the beacon flat using a long average or to send tip-tilt offsets to either the telescope or laboratory adaptive optics.

\subsection{Observing Graphical User Interfaces (GUIs)}
\begin{figure}
\centering
\includegraphics[width=\textwidth]{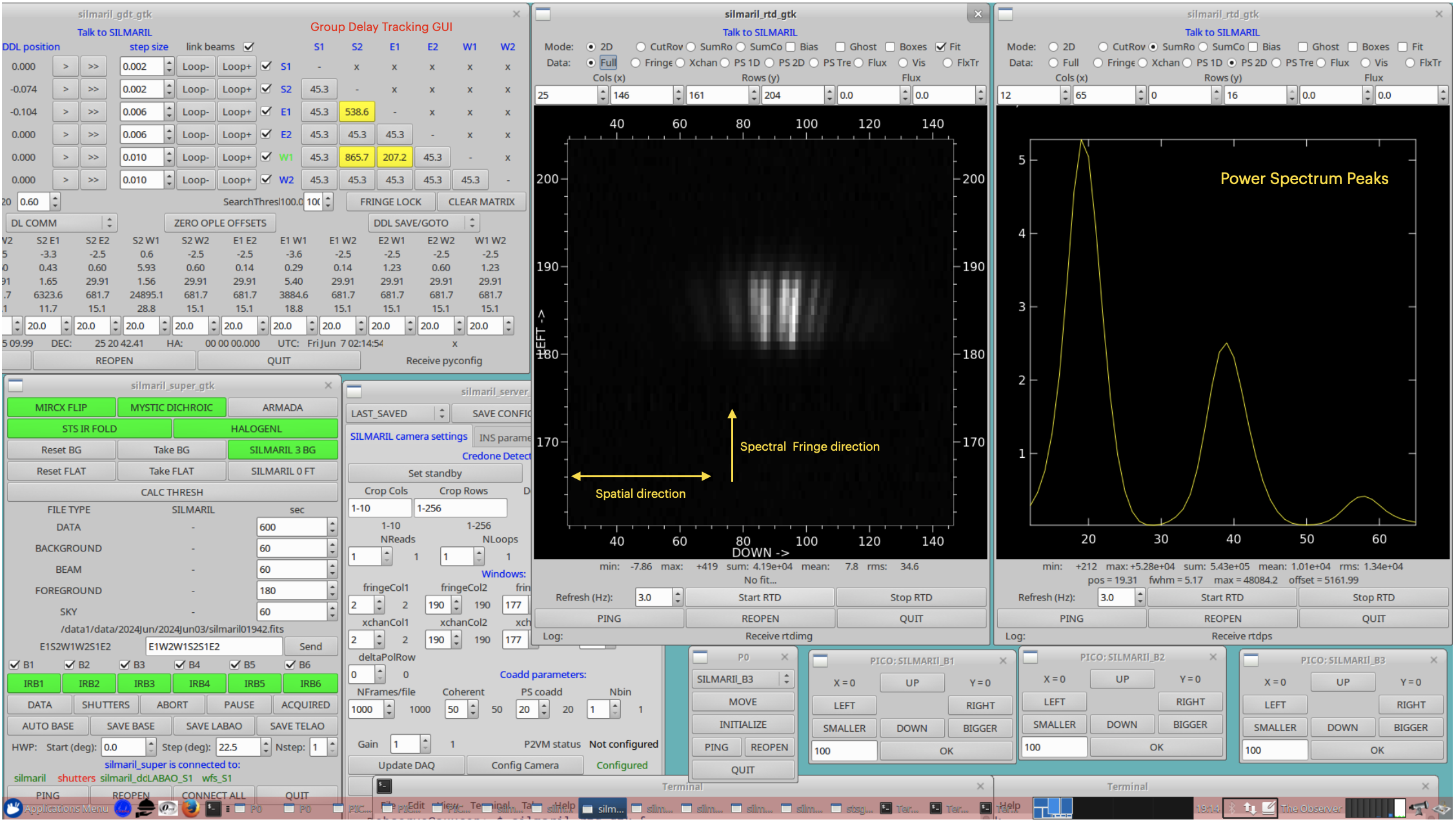}
\caption 
{\label{fig:GUIs} A screenshot of the observing interfaces while taking an observation with the Silmaril instrument. In this observation, the power spectrum peak amplitudes are not equal because of non-uniform illumination. The fringe contrasts were greater than 85\% for all three baseline fringes when uniform fluxes were input to the instrument.  } 
\end{figure}

Silmaril builds upon the standard observing practices established by MIRC-X, which many Principal Investigators are already trained to use. Having similar GUIs to MIRC-X facilitates a quicker transition to the Silmaril observing scheme. To achieve this compatibility, we adapted the MIRC-X observing code and tailored it to the specific requirements of the Silmaril instrument. The GUIs leverage the gtk2 toolkit for their user interface and the plplot library for scientific plotting capabilities. Fig~\ref{fig:GUIs} shows an observing snapshot of the Silmaril instrument. 

\begin{itemize}
    \item \texttt{silmaril\_server\_gtk} -- enables detector and instrument configuration.
    \item \texttt{silmaril\_rtd\_gtk} -- shows real-time display of fringes, power spectrum density, waterfall trends and flux trends, etc. 
    \item \texttt{simaril\_gdt\_gtk} -- enables group delay tracking and locking features
    \item \texttt{silmaril\_ddl\_gtk} -- is an engineering GUI for moving dealy lines
    \item \texttt{picogtk --SILMARIL\_PICO} -- is an engineering GUI for aligning the instrument
\end{itemize}

All computations are implemented within the servers. User actions on the GUIs trigger appropriate function calls inside the servers.

\section{Data Reduction Software}

The Silmaril data reduction pipeline is inspired from \texttt{mircx\_pipeline}\cite{Anugu+2020}.  The Silmaril pipeline is  implemented in \texttt{Python}~3.9, and is publicly available\footnote{\href{https://gitlab.chara.gsu.edu/nanugu/silmaril\_pipeline}{https://gitlab.chara.gsu.edu/nanugu/silmaril\_pipeline}}. The pipeline produces science-ready visibilities ($V^2$) and closure phases ($CP$). Two main differences between the Silmaril and MIRC-X pipelines are:
\begin{itemize}
    \item Beam combination: Silmaril uses a 3-beam combination (resulting in 3 $V^2$ and 1 $CP$) instead of 6-beam combination of MIRC-X/MYSTIC/SPICA (resulting in 15 $V^2$ and 10 $CP$).
    \item Photometry channels: As in Eq.~\ref{Eq.1}, any fluctuations in photometry ($P_{i}$) of different telescopes ($i$), generally caused by variations in atmospheric turbulence, can significantly bias measurements of visibility squared ($V^2$). To address this, beam combiners incorporate dedicated photometric channels alongside their science channels (see ex., MIRC-X/MYSTIC/SPICA). These photometric channels measure the real-time fluxes alongside the fringes, allowing for real-time calibration of $V^2$. However, the Silmaril instrument does not implement these photometric channels on the C-RED1 camera.  To overcome this limitation, we plan to rely on flux measurements from STST and CLIMB to calibrate $V^2$.
\end{itemize}

The data reduction pipeline implements following steps to measure $V^2$ and $CP$:

\begin{itemize}
\item  The raw detector frames undergo background subtraction, bad pixel removal, and flat fielding.
\item Spectral calibration is performed by comparing the observed fringe frequency (in 2.4 px/fringe) to the expected spatial frequency based on the optical magnification. The lateral pupil shifts can affect the fringe frequency, therefore, the spectral calibration is implemented for the small group of  datasets where they were affected.
\item The real-time photometry of each of the 3 beams $P_{i}(\lambda, t)$, where $i$ is  the number of the beam, $t$ is the time (frame), $\lambda$ the spectral channel, are estimated from the photometric channels and the $\kappa$-matrix.
\item  The three coherent fluxes, denoted by $I_{ij}(\lambda, t)$ represent the spatial Fourier components of the interferometric window. These fluxes correspond to the spatial frequencies of each beam pair, where $i$ and $j$ denote the indices of the specific beams involved.
\item Averaged squared visibilities for the baseline $ij$ are computed as follows:
\begin{equation}\label{Eq.1}
    V^2_{ij}(\lambda) = \frac{| \langle \; I_{ij}( \lambda, t) \times I_{ij}(\lambda, t-1)^*  \,-\,\beta(\lambda) \;\rangle |} { 2 \langle P_{i}(\lambda, t) \times P_{j}(\lambda, t) \rangle }
\end{equation}

\item Bispectrum $C$ of closing triangle $ijk$ is calculated by:
\begin{equation}
CP_{ijk}(\lambda)=\langle I_{ij}(\lambda, t) \times I_{jk}(\lambda, t) \times I^{*}_{ik}(\lambda, t)\;-\;\beta_{ijk}(\lambda)\rangle
\end{equation}

\item The instrumental transfer function can be estimated by dividing the observed visibilities of a calibrator source by the expected visibilities for the same source assuming a central uniform disk (UD) model. The UD diameter can be either entered manually or retrieved automatically from a resource like the JMMC Stellar Diameters Catalogue.
\item To account for instrumental variations, transfer function estimates are interpolated to the exact time of each science observation using a Gaussian-weighted average. This average typically considers data within a one-hour window (FWHM = 30 mins) centered on the science observation time. The resulting interpolated transfer function, derived from all calibrators observed with the same instrumental setup, is then used to calibrate the science visibilities.
\end{itemize}

\section{Laboratory Characterization}

To verify the constructed performance of the Silmaril instrument if met its design specifications, we used the laboratory Six Telescope Simulator (STS). We successfully characterized key performance parameters, including the Airy disk size (around 3 pixels at H-band) and large baseline fringe sampling (approximately 2.5 pixels per fringe). These values closely matched to the design specifications. Additionally, fringe contrast across all three baselines exceeded 85\%. Finally, we tested the group delay tracking system by adjusting internal delay lines to achieve co-phasing with the STS, and confirmed its proper functioning. Fig~\ref{fig:cred1} shows the H and K-band fringes, which are meeting the design specifications. We measured the spectral direction is $\sim20$ px in K-band and $\sim12$ px in H-band and those are larger than our design because the prism we use is larger spectral resolution than used in the design.



\section{On-Sky Performance}

\subsection{Slow beam drifts}

Initial on-sky results indicate that we experience slow beam drifts, which can affect the observations in two ways: (i) imperfect beam overlap, leading to biased measurements of fringe contrast and large error bars in $V^2$; and (ii) poor signal-to-noise ratio (SNR), affecting the instrument's sensitivity. To understand the amplitude and frequency of the beam drifts, we collected a large dataset to assess beam drift and long-term stability. The results indicate that we need an active beam stabilization at the order of a few minutes to correct this slow beam drifting. We implement this beam stabilization with the auto alignment software described in Sec.~\ref{Sec:software}.

\subsection{On-Sky Observing Procedure}

The observing procedure starts with selecting the target and configuring the beams to image on the instrument. The first three beams are directed to the left side of the instrument and the last three beams to the right side. The CHARA beam sampler allows for changing the mapping from instrument beam to telescope before the night begins. Telescopes and beams are chosen such that low-frequency baseline fringes of Silmaril are paired with longer telescope baselines, and high-frequency baseline fringes are paired with shorter baselines. As reported in Lanthermann et al. (2022)\cite{Lanthermann2022}, beam 2 \& 3 provide the lower spatial frequency for Silmaril and are paired with longer baseline telescopes, such as S1E1, while beam 1 \& 3 provide the highest spatial frequency and are paired with smaller telescope baselines, for example, S1S2.

\begin{figure}
\centering
\includegraphics[width=0.8\textwidth]{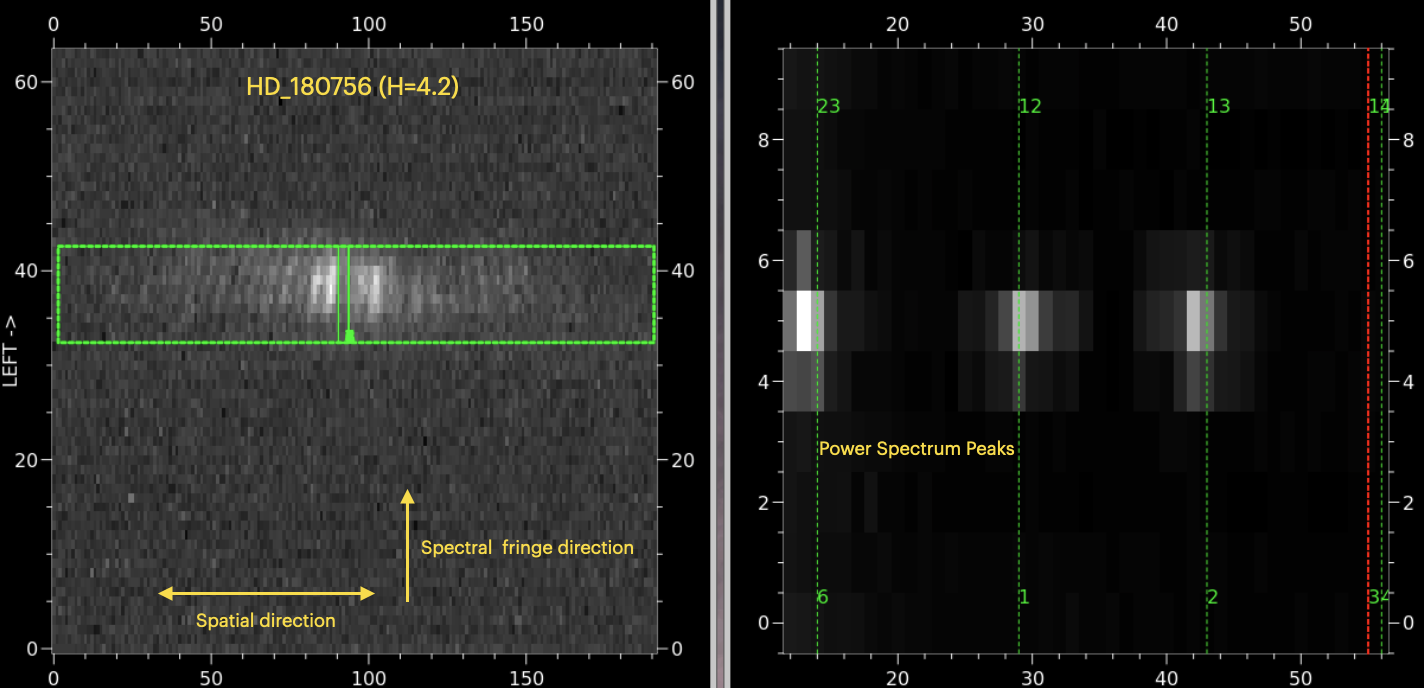}
\caption 
{\label{fig:cred2_on_sky} A snapshot of an on-sky observation of HD 180756 displaying fringes and the power spectrum. This observation was obtained with the C-RED2 camera.} 
\end{figure}

Before the night, the alignment of Silmaril is optimized using the Six Telescope Simulator (STS) , which is the 6-beam coherent calibration source at the CHARA Array. The goal is to ensure that the light is well-centered within the predefined coordinates on the Silmaril detector and cophased with the STS, helping to find the fringes within ±1 mm offsets by the astrometric solution. While the instrument calibration is performed daily, it has proven stable for at least a week unless the observing mode has changed. First, the beams are centered on the predefined reference pixels vertically and horizontally. Next, the power spectrum fringe peaks are checked to ensure they fall onto their predetermined positions (see Fig~\ref{fig:cred2_on_sky}). This calibration is accomplished by adjusting the cylindrical mirrors (CM1s).

The star is initially acquired by the CHARA Adaptive Optics (AO) system, which operates at visible wavelengths. Visible to near-infrared, atmospheric refraction star shifts are corrected by positioning the star within the reference boxes on the J+H-band operating the STST camera. This correction is implemented by adjusting the beacon flat on the telescope area to center the beams on Silmaril. These STST reference boxes are defined by the STS laboratory white light source.

At this stage, the Silmaril instrument should see light on the detector. If any misalignments between Silmaril and STST, those are corrected with a coarse alignment by adjusting the beacon flat on the telescope area to center the beams on Silmaril. Once beam stabilization is activated, fringes are typically found within 5 minutes of observing time, depending on the CHARA astrometric baseline, which is calibrated quarterly following seasonal trends. The data acquisition sequence is similar to other instruments.

The standard observing sequence is as follows, with 20 mins for an observation. However, for fainter targets, longer integrations times of the fringe DATA are necessary. 

\begin{itemize}
    \item Star Acquisition: 5 minutes to acquire the star with the telescopes.

    \item Fringe Search: 2–5 minutes of searching for fringes.

    \item Data Collection: 5 minutes of DATA collection with all shutters open and fringes tracked.

    \item Background Measurement: 1 minute with all shutters closed.

    \item Beam Measurement: 1 minute for each BEAM, where the shutter of only one of the six beams is open sequentially.

    \item Foreground Frames: 3 minutes of data without fringes with all shutters open. Delay lines are moved away while taking this dataset.

    \item Sky Measurement: 1 minute by moving the telescopes away off the target.
\end{itemize}

\subsection{Results from Engineering Camera C-RED2}
\begin{figure}
\centering
\includegraphics[width=0.49\textwidth]{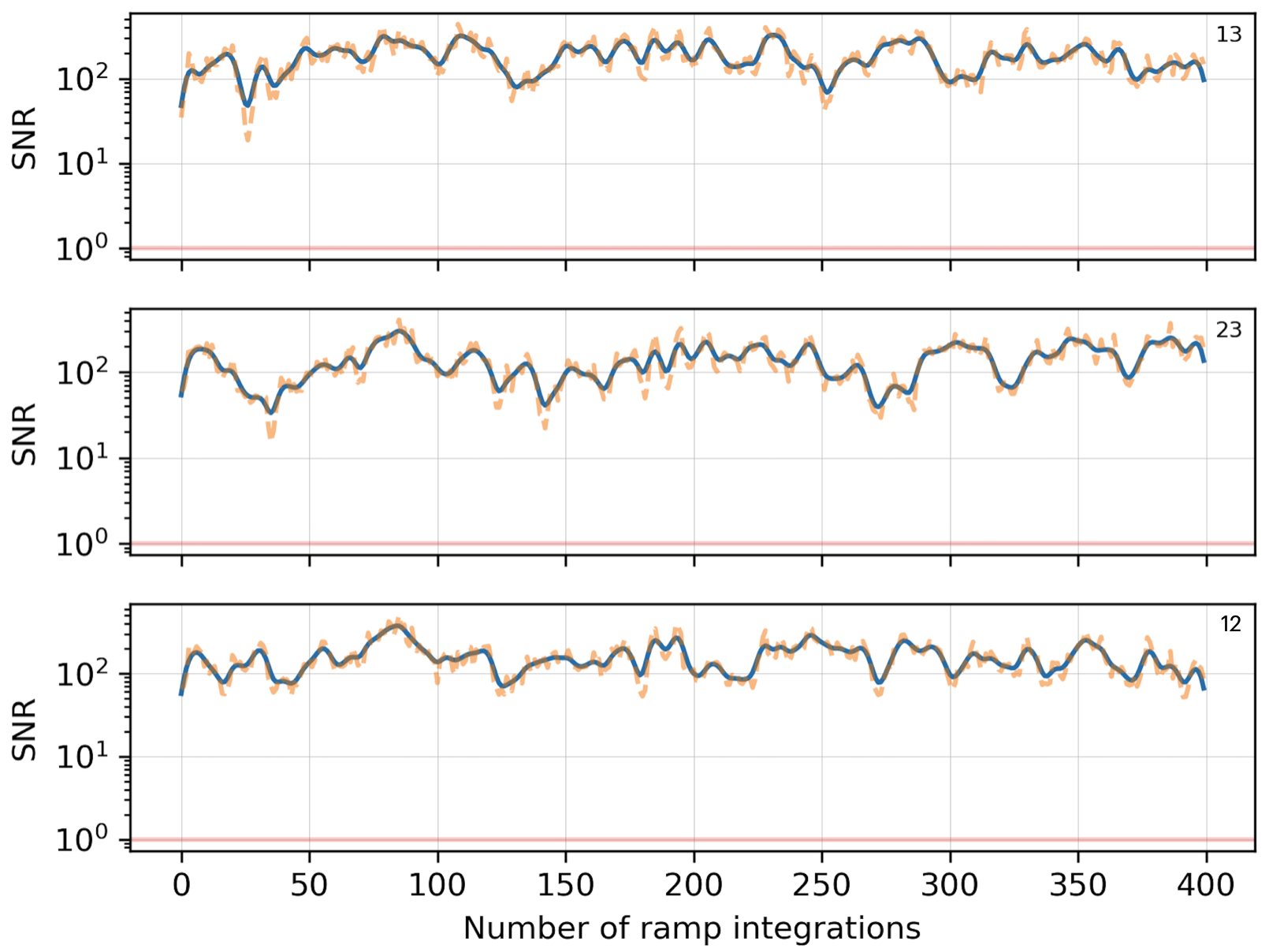}
\includegraphics[width=0.49\textwidth]{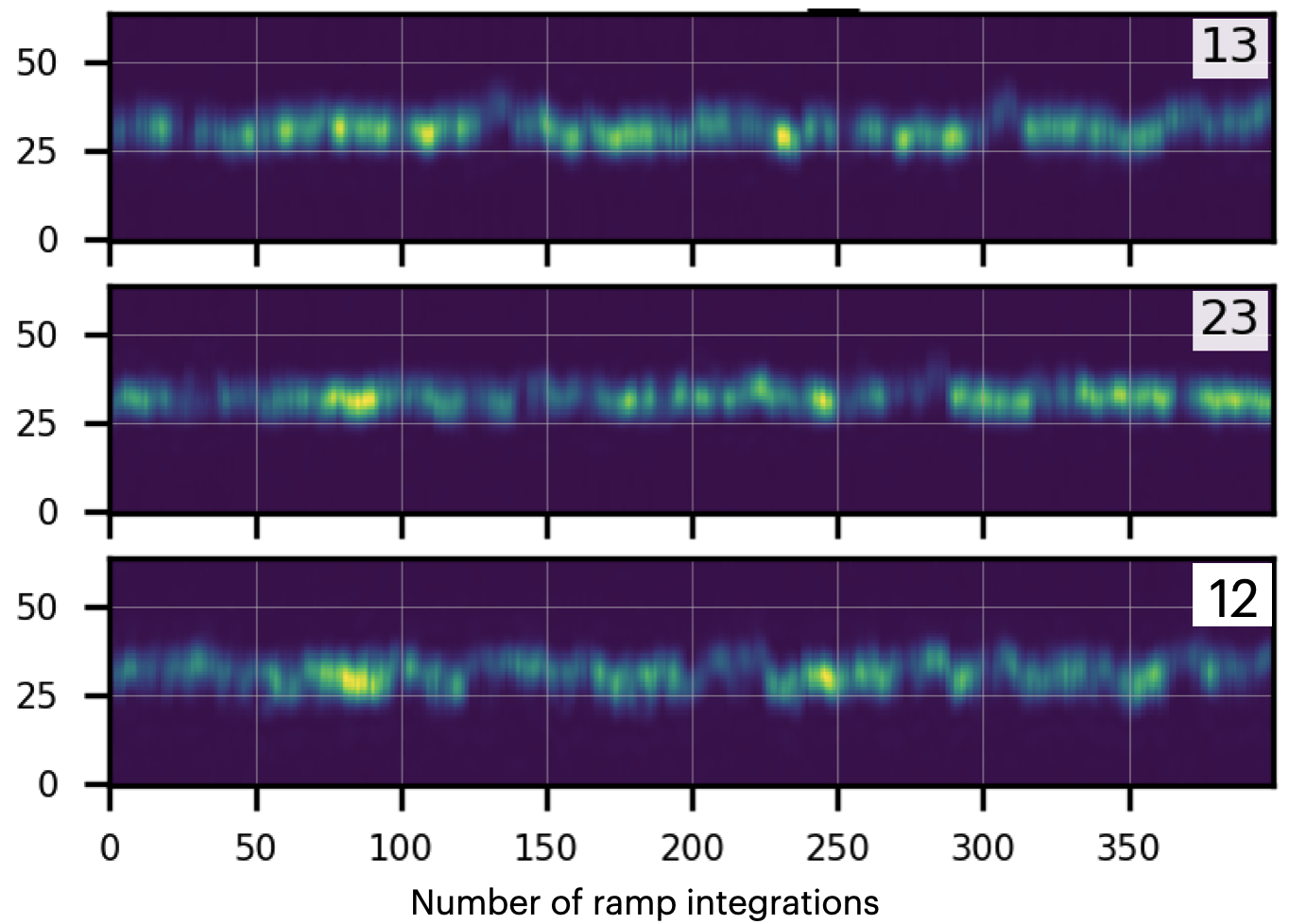}
\caption 
{\label{fig:SNR} (Left panel) The fringe signal-to-noise ratio ($\sim100$) of an observation recorded on the object HD 180756 with an H magnitude of 4.2 was recorded with C-RED2. The bottom orange line indicates the signal-to-noise ratio (SNR=2) threshold for fringe rejection. Any fringe with a signal falling below this threshold will be rejected. (Right panel) The waterfall trends for three baselines  recorded on the object HD 180756.  This observation is projected to be around 
$H\gtrapprox 10$ if observed with C-RED1, considering the ultra-low noise performance of the C-RED1 detector compared to C-RED2. } 
\end{figure}

While awaiting the readiness of the science-grade C-RED1 camera, we proceeded with on-sky observations using the engineering camera C-RED2. However, the sensitivity of C-RED2 is limited compared to C-RED1 due to the absence of key features such as avalanche gain for reducing readout noise, lowered dark current, and a cold stop to minimize black body backgrounds. Nonetheless, with C-RED2, we successfully characterized the instrument: (i) obtained on-sky fringes, (ii) demonstrated on-sky group delay fringe tracking, and (iii) collected data on several stars to test the data reduction pipeline. We successfully captured high SNR fringes on HD~180756 (magnitude 4.2 in H band) with a 50 ms integration. Fig \ref{fig:SNR}  illustrates results from the data reduction pipeline, showing the signal-to-noise ratio of the observations. 

Considering the ultra low noise capabilities of C-RED1, we compute that Silmaril will be able to detect much fainter sources ($H/K \gtrapprox 10$) once we transition from C-RED2 to C-RED1. We are looking forward to the arrival of C-RED1 this summer, which will allow us to test the final sensitivity limits of the Silmaril instrument. Please refer to Lanthermann et al. (2024) \cite{Lanthermann2024} for more details on the sensitivity calculations.

\section{Summary and Future Prospects}
Silmaril, a new beam combiner for the CHARA array, promises access to previously unseen faint near-infrared targets (magnitude   $H/K \gtrapprox 10$). This paper outlines the instrument's software suite and data reduction pipeline, responsible for camera control, fringe tracking, instrument alignment, data processing, and user interfaces. 

Future Prospects:
\begin{itemize}
\item On-sky Testing of Science Ready Camera: While initial on-sky observations have been conducted using the C-RED2 engineering camera, simulations based on C-RED1's ultra-low noise characteristics predict Silmaril achieving sensitivity $H \gtrapprox 10$. This enhanced capability will be tested upon the repaired C-RED1 camera's return to CHARA in July 2024.
\item Improved Beam Stability: Initial on-sky observations revealed slow beam drifts challenges. To address this issue, we have implemented automatic alignment methods that should stabilize the beams.
\item Data Reduction Improvement: The data reduction software is still under refinement to improve the accuracy of squared visibility and closure phase measurements. Additionally, the software needs refinement to calculate uncertainties more accurately, as current values are underestimated.
\item Community Access: We anticipate that Silmaril will be ready for CHARA community access observations starting 2025A semester.
\end{itemize}

\subsection* {Acknowledgments}
The construction of the Silmaril instrument was supported by USA National Science Foundation, under Grant No. NSF AST-1909858. \\
This work is based upon observations obtained with the Georgia State University 
Center for High Angular Resolution Astronomy Array at Mount Wilson Observatory.  
The CHARA Array is supported by the National Science Foundation under Grant No.\
AST-1636624, AST-1908026, and AST-2034336.  Institutional support has been provided 
from the GSU College of Arts and Sciences and the GSU Office of the Vice President 
for Research and Economic Development.


\bibliography{report}   

\begin{thebibliography}{1}

\bibitem{tenBrummelaar+2005}
{ten Brummelaar}, T.~A., {McAlister}, H.~A., {Ridgway}, S.~T., {Bagnuolo}, W.~G., J., {Turner}, N.~H., {Sturmann}, L., {Sturmann}, J., {Berger}, D.~H., {Ogden}, C.~E., {Cadman}, R., {Hartkopf}, W.~I., {Hopper}, C.~H., and {Shure}, M.~A., ``{First Results from the CHARA Array. II. A Description of the Instrument},'' {\em Astrophysical Journal}~{\bf 628},  453--465 (July 2005).

\bibitem{Lanthermann2022}
{Lanthermann}, C., {ten Brummelaar}, T., {Tuthill}, P., {Martinod}, M.-A., {Ligon}, E., {Gies}, D., {Schaefer}, G., and {Anderson}, M., ``{Design of the new CHARA instrument SILMARIL: pushing for the sensitivity of a 3-beam combiner in the H- and K-bands},'' in [{\em Optical and Infrared Interferometry and Imaging VIII}{\nolinebreak\hspace{0.1em}]},  {M{\'e}rand}, A., {Sallum}, S., and {Sanchez-Bermudez}, J., eds., {\em Society of Photo-Optical Instrumentation Engineers (SPIE) Conference Series} {\bf 12183},  121830N (Aug. 2022).

\bibitem{Lanthermann2024}
{Lanthermann}, C., {ten Brummelaar}, T., {Tuthill}, P., {Anugu}, N., {Ligon}, E., {Gies}, D., {Schaefer}, G., and {Anderson}, M., ``{Silmaril: final design and on-sky performance},'' in [{\em Optical and Infrared Interferometry and Imaging IX}{\nolinebreak\hspace{0.1em}]},  {M{\'e}rand}, A., {Sallum}, S., and {Sanchez-Bermudez}, J., eds., {\em Society of Photo-Optical Instrumentation Engineers (SPIE) Conference Series} {\bf 13095-4} (Aug. 2024).

\bibitem{Mourard+2022}
{Mourard}, D., {Berio}, P., {Pannetier}, C., {Nardetto}, N., {Albrecht}, S., {Allouche}, F., {Bailet}, C., {Bourgès}, L., {ten Brummelaar}, Theo A.and~{Creevey}, O., {Deheuvels}, S., {Dejonghe}, J., {Domiciano}, A., {Geneslay}, P., {Gies}, D.~R., {Jacqmart}, E., {Lagarde}, S., {Lecron}, D., {Ligi}, R., {Mella}, G., {Morand}, F., {Rousseau}, S., {Salabert}, D., {Schaefer}, G., {Spang}, A., and {Wittkowski}, M., ``{CHARA/SPICA: a new 6T instrument for the CHARA Array},'' in [{\em Optical and Infrared Interferometry and Imaging VIII}{\nolinebreak\hspace{0.1em}]},  {M\'erand}, A., {Sallum}, S., and {Sanchez-Bermudez}, J., eds., {\em Society of Photo-Optical Instrumentation Engineers (SPIE) Conference Series} {\bf 12183},  12183--7 (July 2022).

\bibitem{Anugu+2020}
{Anugu}, N., {Le Bouquin}, J.-B., {Monnier}, J.~D., {Kraus}, S., {Setterholm}, B.~R., {Labdon}, A., {Davies}, C.~L., {Lanthermann}, C., {Gardner}, T., {Ennis}, J., {Johnson}, K. J.~C., {Ten Brummelaar}, T., {Schaefer}, G., and {Sturmann}, J., ``{MIRC-X: A Highly Sensitive Six-telescope Interferometric Imager at the CHARA Array},'' {\em Astronomical Journal}~{\bf 160},  158 (Oct. 2020).

\bibitem{Setterholm2023JATIS...9b5006S}
{Setterholm}, B.~R., {Monnier}, J.~D., {Le Bouquin}, J.-B., {Anugu}, N., {Ennis}, J., {Jocou}, L., {Ibrahim}, N., {Kraus}, S., {Anderson}, M.~D., {Chhabra}, S., {Codron}, I., {Farrington}, C.~D., {Flores}, B., {Gardner}, T., {Gutierrez}, M., {Lanthermann}, C., {Majoinen}, O.~W., {Mortimer}, D.~J., {Schaefer}, G., {Scott}, N.~J., {ten Brummelaar}, T., and {Vargas}, N.~L., ``{MYSTIC: a high angular resolution K-band imager at CHARA},'' {\em Journal of Astronomical Telescopes, Instruments, and Systems}~{\bf 9},  025006 (Apr. 2023).

\bibitem{Taras2024ApOpt..63D..41T}
{Taras}, A.~K., {Robertson}, J.~G., {Allouche}, F., {Courtney-Barrer}, B., {Carter}, J., {Crous}, F., {Cvetojevic}, N., {Ireland}, M., {Lagarde}, S., {Martinache}, F., {McGinness}, G., {N'Diaye}, M., {Robbe-Dubois}, S., and {Tuthill}, P., ``{Heimdallr, Baldr, and Solarstein: designing the next generation of VLTI instruments in the Asgard suite},'' {\em \ao}~{\bf 63},  D41 (May 2024).

\bibitem{tenBrummelaar+2013}
{Ten Brummelaar}, T.~A., {Sturmann}, J., {Ridgway}, S.~T., {Sturmann}, L., {Turner}, N.~H., {McAlister}, H.~A., {Farrington}, C.~D., {Beckmann}, U., {Weigelt}, G., and {Shure}, M., ``{The Classic/climb Beam Combiner at the CHARA Array},'' {\em Journal of Astronomical Instrumentation}~{\bf 2},  1340004 (Dec. 2013).

\end{thebibliography}
\bibliographystyle{spiebib}   

\end{document}